\documentclass[twocolumn,showpacs,preprintnumbers,amsmath,amssymb]{revtex4}

\usepackage{graphicx}
\usepackage{dcolumn}
\usepackage{bm}

\begin{document}

\preprint{}

\title{Real-Valued Charged Fields\\ and Interpretation of Quantum Mechanics II}

\author{Andrey Akhmeteli}
 \email{akhmeteli@aim.com}
\affiliation{%
Intelligent Optical Systems, Inc.\\
2520 W. 237th Street\\
Torrance, CA 90505, USA}%

 \homepage{http://www.akhmeteli.org}

\date{\today}

\begin{abstract}
In the first part of this work (http://www.arxiv.org/abs/quant-ph/0509044), it was shown that the Klein-Gordon-Maxwell electrodynamics in the unitary gauge allows natural elimination of the particle wave function and describes independent evolution of the electromagnetic field. Therefore, the electromagnetic field can be regarded as the guiding field in the  Bohm interpretation of quantum mechanics. An extension of those results to the Dirac-Maxwell electrodynamics was less general, but represented at least an interesting toy model of quantum theory. Another model based on the Dirac-Maxwell electrodynamics is considered in this work. The model also typically allows elimination of the wave function and describes independent evolution of the electromagnetic field.
\end{abstract}

\maketitle

\section{\label{sec:level1}Introduction}

It was shown in the first part of this work (Ref.~\cite{Akhm10}) that the Klein-Gordon-Maxwell electrodynamics in the unitary gauge, where the wave function is real, allows natural elimination of the particle wave function and describes independent evolution of the electromagnetic field in the following sense: if the components of the electromagnetic potential vector and their first derivatives with respect to time are known in the entire space at some point in time, their second derivatives with respect to time can be calculated from the equations of motion, so integration yields these components for any point in time. The particle current vector has the same direction as the electromagnetic potential vector, so the latter fully determines the Bohmian trajectories and may be regarded as the guiding field in the  Bohm (de Broglie-Bohm) interpretation of quantum mechanics. These results can be extended to the case where conserved external currents are present. Such an extension can be useful, e.g., for a discussion of the Aharonov-Bohm effect.

The results of Ref.~\cite{Akhm10} for the Dirac-Maxwell electrodynamics were less general, as the equations of motion for the relevant model were obtained by imposition of the constraint $\bar{\Psi}\gamma^5\gamma^\mu\Psi=0$ (the axial current vanishes) for the standard Lagrangian of the Dirac-Maxwell electrodynamics, so the equations differed from the standard ones. It is not clear whether the equations are compatible with experimental data, but they certainly can be used as an interesting toy model for interpretation of quantum mechanics, as they also allow natural elimination of the particle wave function and describe independent evolution of the electromagnetic field. In this second part of the work, a modified version of this model is considered. The equations of motion for this modified version could be formally obtained by imposition of the constraint $\imath\bar{\Psi}\gamma^5\gamma^\mu\Psi=0$ (the axial current times $\imath$ vanishes) for the standard Lagrangian of the Dirac-Maxwell electrodynamics. As this condition is not real, such a procedure cannot be justified. However, the resulting model may be of some interest, as the relevant equations have a relatively rich set of solutions. The model also typically allows natural elimination of the particle wave function and describes independent evolution of the electromagnetic field. Furthermore, the existence of these properties is possible due to a nontrivial mechanism, which can be useful for other models.
 
It should be emphasized that solutions of partial differential equations are only considered locally in both parts of this work, although global properties of the solutions can be important for physics.

\maketitle

\section{\label{sec:level1}Majorana solutions of the modified Dirac-Maxwell electrodynamics}

Let us start with the following equations of motion (again, they cannot be properly justified, but have some interesting properties):
\begin{equation}\label{eq:pr25}
(i\hat{\partial}-e\hat{A}+\imath\gamma^5\hat{D}-m)\Psi=0,
\end{equation}
\begin{equation}\label{eq:pr26}
\Box A_\mu-A^\nu_{,\nu\mu}=j_\mu,
\end{equation}
\begin{equation}\label{eq:pr27}
j_\mu=e\bar{\Psi}\gamma_\mu\Psi,
\end{equation}
\begin{equation}\label{eq:pr28}
j_a^\mu=\bar{\Psi}\gamma^5\gamma^\mu\Psi=0,
\end{equation}
where $\hat{D}=D_\mu\gamma^\mu$, and $D_\mu$ are the Lagrangian multipliers (they can be regarded as some ghost fields). Every solution of this system is physically equivalent to a Majorana solution related to it via a gauge transform: Eq.~(\ref{eq:pr28}) implies that the spinor $\Psi$ may be represented in the form $\Psi=\exp(i\theta)\Phi$, where $\theta=\theta(x)$ is real, and $\Phi$ is a spinor satisfying the Majorana condition. Let us prove this statement. In the Majorana representation (see Refs.~\cite{Itzykson},~\cite{Akhm10}), the components of $j_a^\mu$ are:
\begin{eqnarray}
j_a^0=2 \mathrm{Im}(\Psi_1\Psi_2^*+\Psi_4\Psi_3^*)=0,
\nonumber\\
j_a^1=2 \mathrm{Im}(\Psi_3\Psi_1^*+\Psi_2\Psi_4^*)=0,
\nonumber\\
j_a^2=2 \mathrm{Im}(\Psi_1\Psi_2^*+\Psi_3\Psi_4^*)=0,
\nonumber\\
j_a^3=2 \mathrm{Im}(\Psi_1\Psi_4^*+\Psi_2\Psi_3^*)=0.
\nonumber
\end{eqnarray}
We obtain from the expressions for $j_a^0$ and $j_a^2$ that $\Psi_1\Psi_2^*$ and $\Psi_4\Psi_3^*$ are real, so $\Psi_\alpha$ can be presented in the following form:
\begin{eqnarray}
\Psi_1=r_1 \exp(\imath\theta),\Psi_2=r_2 \exp(\imath\theta),
\nonumber\\
\Psi_3=r_3 \exp(\imath\beta),\Psi_4=r_4 \exp(\imath\beta),
\nonumber
\end{eqnarray}
where $r_\alpha$, $\theta$, and $\beta$ are real.
We obtain from the expressions for $j_a^1$ and $j_a^3$ that
\begin{eqnarray}
r_3 r_1\sin(\beta-\theta)+r_2 r_4\sin(\theta-\beta)=0,
\nonumber\\
r_1 r_4\sin(\theta-\beta)+r_2 r_3\sin(\theta-\beta)=0.
\nonumber
\end{eqnarray}
If $\sin(\theta-\beta)=0$, the statement can be easily proven, if not, then
\begin{equation}
r_3 r_1-r_2 r_4=r_1 r_4+r_2 r_3=0,
\end{equation}
therefore, $r_2(r_3^2+r_4^2)=0$. If $r_3^2+r_4^2=0$, the statement can be easily proven, otherwise $r_2=0$ and $r_3 r_1=r_1 r_4=0$, so either $r_1=0$ or $r_3=r_4=0$. In either case the statement can be easily proven.
 Substituting $\Psi=\exp(i\theta)\Phi$ in Eqs.~(\ref{eq:pr25},\ref{eq:pr26},\ref{eq:pr27}), we obtain equations for Majorana spinors:
\begin{equation}\label{eq:pr29}
(i\hat{\partial}-e\hat{B}+\imath\gamma^5\hat{D}-m)\Phi=0,
\end{equation}
\begin{equation}\label{eq:pr30}
\Box B_\mu-B^\nu_{,\nu\mu}=j_\mu,
\end{equation}
\begin{equation}\label{eq:pr31}
j_\mu=e\bar{\Phi}\gamma_\mu\Phi,
\end{equation}
where $e B_\mu=e A_\mu+\theta_{,\mu}$. Applying charge conjugation to Eq.~(\ref{eq:pr29}) and using the Majorana condition, we obtain:
\begin{equation}\label{eq:pr32}
(i\hat{\partial}-m)\Phi=0,
\end{equation}
\begin{equation}\label{eq:pr33}
(-e\hat{B}+\imath\gamma^5\hat{D})\Phi=0.
\end{equation}
If $\Phi(x)\neq 0$, one can use linear algebra to eliminate $D_\mu$ from Eq.~(\ref{eq:pr33}) and obtain for point $x$
\begin{equation}\label{eq:pr33a1}
j_\mu B^\mu=0.
\end{equation}
Actually, to obtain Eq.~(\ref{eq:pr33a1}), one can multiply Eq.~(\ref{eq:pr33}) by $\bar{\Phi}$ from the left and use Eq.~(\ref{eq:pr28}). Further analysis shows that if Eq.~(\ref{eq:pr33a1}) is satisfied, there exist such $D_\mu$ that Eq.~(\ref{eq:pr33}) is satisfied. Indeed, let us use a fixed frame of reference, and introduce for each point $x$ two real vectors $V$ and $U$ defined by the following coordinates:
\begin{eqnarray}\label{eq:pr33a2}
V^\mu=V^\mu(x)=\imath\bar\Phi(x)\gamma^0\gamma^\mu\Phi(x),
\nonumber\\
U^\mu=U^\mu(x)=\bar\Phi(x)\gamma^0\gamma^5\gamma^\mu\Phi(x).
\end{eqnarray}
Then it can be proven by a straight-forward calculation that if vector $B$ satisfies Eq.~(\ref{eq:pr33a1}), then Eq.~(\ref{eq:pr33}) is satisfied by the following vector:
\begin{equation}\label{eq:pr33a3}
D=-e\frac{(B_\mu V^\mu)U-(B_\mu U^\mu)V}{V_\mu V^\mu}.
\end{equation}

Let us consider the system of Eqs.~(\ref{eq:pr30},\ref{eq:pr31},\ref{eq:pr32},\ref{eq:pr33a1}). One can prove directly that the system has a rather broad set of solutions. Indeed, we can choose values of the Majorana spinor $\Phi$ arbitrarily on the hyperplane $x^0=0$. The free Dirac equation Eq.~(\ref{eq:pr32}) then determines the spinor for all values of $x^0$. We can then eliminate $B^0$ and $\dot{B^0}$ from Eq.~(\ref{eq:pr30}) using Eq.~(\ref{eq:pr33a1}) and its derivative with respect to $x^0$. Then Eq.~(\ref{eq:pr30}) for $\mu=0$ can be regarded as a constraint on $B^k$ and $\dot{B^k}$, where $k=1,2,3$. If $B^k$ and $\dot{B^k}$ satisfy this constraint for $x^0=0$, but are otherwise chosen arbitrarily on this hyperplane (for example, we can choose $B^1$, $B^2$, $B^3$, $\dot{B^1}$, $\dot{B^2}$ arbitrarily for $x^0=0$ and $\dot{B^3}$ arbitrarily for $x^0=x^3=0$), Eq.~(\ref{eq:pr30}) for $\mu=1,2,3$ will yield a solution of Eq.~(\ref{eq:pr30}) for all values of $x^0$, as the constraint will be satisfied for all $x^0$, because its derivative with respect to $x^0$ vanishes due to conservation of the current for the Dirac equation.

It is evident that the system of Eqs.~(\ref{eq:pr30},\ref{eq:pr31},\ref{eq:pr32},\ref{eq:pr33a1}) describes independent evolution of the Majorana spinor field (see Eq.~(\ref{eq:pr32})). It is less obvious that it also typically describes independent evolution of the electromagnetic field $B^\mu$. Specifically, it is possible to prove that if $B^\mu$, $\dot{B^\mu}$, $\ddot{B^\mu}$, and $B^{\mu(III)}$ are known for a certain value of $x^0$ in the entire hyperplane defined by this value, $B^{\mu(IV)}$ can be typically found from the system for the same hyperplane (here $B^{\mu(III)}$ and $B^{\mu(IV)}$ are the third and the fourth derivatives of $B^\mu$ with respect to $x^0$). Thus, the Cauchy problem can be posed and solved for all $x^0$.

To prove this, let us first establish what can be said about a Majorana spinor $\Phi$, if we know the relevant current $J^\mu=\bar{\Phi}\gamma^\mu\Phi$. A straight-forward calculation shows that spinor
\begin{equation}\label{eq:pr33a4}
\psi=\frac{1}{\sqrt{2(J^0+J^2)}}\left(
        \begin{array}{c}
          0 \\
          J^0+J^2 \\
          -J^1 \\
          J^3 \\
        \end{array}
      \right)
\end{equation}
has the same current.

Let us now find a relation between two Majorana spinors $\Phi$ and $\psi$ that have the same current. If we introduce two complex variables $\xi^1=\Phi^1+\imath\Phi^2$ and $\xi^2=\Phi^3+\imath\Phi^4$ for spinor $\Phi$ and corresponding variables $\xi'^{1}$ and $\xi'^{2}$ for spinor $\psi$, then, for example,
\begin{eqnarray}\label{eq:pr33a5}
\xi^1\xi^{1*}=\frac{1}{2}(J^0+J^2),
\nonumber\\
\xi^2\xi^{2*}=\frac{1}{2}(J^0-J^2),
\nonumber\\
\xi^1\xi^{2}=-\frac{1}{2}(J^3+\imath J^1).
\end{eqnarray}
A straight-forward calculation shows that it is possible to find such real value $\phi$ that
\begin{eqnarray}\label{eq:pr33a6}
\xi^1=\xi'^1\exp(-\imath\phi),
\nonumber\\
\xi^2=\xi'^2\exp(\imath\phi),
\end{eqnarray}
which translates for $\Phi$ and $\psi$ into
\begin{equation}\label{eq:pr33a7}
\Phi=\exp(\imath\gamma^5\phi)\psi.
\end{equation}
Thus, if the current is the same for Majorana spinors $\Phi$ and $\psi$, there is a relationship Eq.~(\ref{eq:pr33a7}) between them. Let us substitute Eq.~(\ref{eq:pr33a7}) in the free Dirac equation Eq.~(\ref{eq:pr32}):
\begin{eqnarray}\label{eq:pr33a8}
0=\imath\gamma^\mu\partial_\mu(\exp(\imath\gamma^5\phi)\psi)
-m\exp(\imath\gamma^5\phi)\psi=
\nonumber\\
=\imath\gamma^\mu(\imath\gamma^5\phi_{,\mu}\exp(\imath\gamma^5\phi)\psi+
\exp(\imath\gamma^5\phi)\psi_{,\mu})-
\nonumber\\
-m\exp(\imath\gamma^5\phi)\psi=
\gamma^5\exp(-\imath\gamma^5\phi)\gamma^\mu\phi_{,\mu}\psi+
\nonumber\\
+\imath\exp(-\imath\gamma^5\phi)\gamma^\mu\psi_{,\mu}
-m\exp(\imath\gamma^5\phi)\psi,
\end{eqnarray}
or
\begin{equation}\label{eq:pr33a9}
\gamma^5\gamma^\mu\phi_{,\mu}\psi+\imath\gamma^\mu\psi_{,\mu}-
m\exp(2\imath\gamma^5\phi)\psi=0.
\end{equation}
If we multiply Eq.~(\ref{eq:pr33a9}) by $\bar{\psi}\gamma^0$ from the left, we obtain
\begin{equation}\label{eq:pr33a10}
\phi_{,\mu}\bar{\psi}\gamma^0\gamma^5\gamma^\mu\psi+
\imath\bar{\psi}\gamma^0\gamma^\mu\psi_{,\mu}-
m\cos(2\phi)\bar{\psi}\gamma^0\psi=0,
\end{equation}
as the axial current $\bar{\psi}\gamma^5\gamma^\mu\psi$ vanishes for Majorana spinors, so
\begin{eqnarray}\label{eq:pr33a11}
\bar{\psi}\gamma^0\exp(2\imath\gamma^5\phi)\psi=
\bar{\psi}\gamma^0(\cos(2\phi)+\imath\gamma^5\sin(2\phi))\psi=
\nonumber\\
=\cos(2\phi)\bar{\psi}\gamma^0\psi.
\end{eqnarray}
If we multiply Eq.~(\ref{eq:pr33a9}) by $\imath\bar{\psi}\gamma^0\gamma^5$ from the left, we obtain in the same way:
\begin{equation}\label{eq:pr33a12}
\phi_{,\mu}\imath\bar{\psi}\gamma^0\gamma^\mu\psi-
\bar{\psi}\gamma^0\gamma^5\gamma^\mu\psi_{,\mu}+
m\sin(2\phi)\bar{\psi}\gamma^0\psi=0.
\end{equation}
For the same fixed frame of reference, let us introduce for each point $x$ two real vectors $v=v(x)$ and $u=u(x)$ and two real scalars $q=q(x)$ and $r=r(x)$ defined as follows:
\begin{eqnarray}\label{eq:pr33a13}
v^\mu=\frac{\imath\bar{\psi}\gamma^0\gamma^\mu\psi}{\bar{\psi}\gamma^0\psi},
\nonumber\\
u^\mu=\frac{\bar{\psi}\gamma^0\gamma^5\gamma^\mu\psi}{\bar{\psi}\gamma^0\psi},
\nonumber\\
q=\frac{\bar{\psi}\gamma^0\gamma^5\gamma^\mu\psi_{,\mu}}{\bar{\psi}\gamma^0\psi},
\nonumber\\
r=\frac{-\imath\bar{\psi}\gamma^0\gamma^\mu\psi_{,\mu}}{\bar{\psi}\gamma^0\psi}.
\end{eqnarray}
Eqs.~(\ref{eq:pr33a10},\ref{eq:pr33a12}) can then be rewritten as follows:
\begin{eqnarray}\label{eq:pr33a14}
v^\mu\phi_{,\mu}=q-m\sin(2\phi),
\nonumber\\
u^\mu\phi_{,\mu}=r+m\cos(2\phi).
\end{eqnarray}
A straight-forward calculation shows the following for a Majorana spinor $\psi$:
\begin{eqnarray}\label{eq:pr33a15}
v^0=u^0=0,
\\
v^\mu v_\mu=u^\mu u_\mu=-1,
\\
v^\mu u_\mu=0.
\end{eqnarray}
Let us apply the commutator of Lie derivatives $v^\mu \partial_\mu$ and  $u^\mu \partial_\mu$ to $\phi$, using Eq.~(\ref{eq:pr33a14}):
\begin{eqnarray}\label{eq:pr33a16}
u^\nu\partial_\nu(v^\mu\phi_{,\mu})-v^\nu\partial_\nu(u^\mu\phi_{,\mu})=
(u^\nu v^\mu_{,\nu}-v^\nu u^\mu_{,\nu})\phi_{,\mu}=
\nonumber\\
=u^\nu\partial_\nu(q-m\sin(2\phi))-v^\nu\partial_\nu(r+m\cos(2\phi))=
\nonumber\\
=u^\nu q_{,\nu}-2m\cos(2\phi)u^\nu\phi_{,\nu}-v^\nu r_{,\nu}+
\nonumber\\
+2m\sin(2\phi)v^\nu\phi_{,\nu}=u^\nu q_{,\nu}-v^\nu r_{,\nu}-
\nonumber\\
-2m\cos(2\phi)(r+m\cos(2\phi))+
\nonumber\\
+2m\sin(2\phi)(q-m\sin(2\phi))=u^\nu q_{,\nu}-v^\nu r_{,\nu}-
\nonumber\\
-2m r\cos(2\phi)+2m q\sin(2\phi)-2m^2.
\end{eqnarray}
Let us define vector $w$ by its coordinates
\begin{equation}\label{eq:pr33a17}
w^\mu=u^\nu v^\mu_{,\nu}-v^\nu u^\mu_{,\nu}+2r u^\mu+2q v^\mu,
\end{equation}
then, from Eqs.~(\ref{eq:pr33a14},\ref{eq:pr33a16}), we obtain
\begin{equation}\label{eq:pr33a18}
w^\mu\phi_{,\mu}=u^\nu q_{,\nu}-v^\nu r_{,\nu}+2r^2+2q^2-2m^2,
\end{equation}
or
\begin{equation}\label{eq:pr33a19}
w^\mu\phi_{,\mu}=p,
\end{equation}
where
\begin{equation}\label{eq:pr33a20}
p=u^\nu q_{,\nu}-v^\nu r_{,\nu}+2r^2+2q^2-2m^2.
\end{equation}
Let us apply the commutator of Lie derivatives $v^\mu \partial_\mu$ and  $w^\mu \partial_\mu$ to $\phi$:
\begin{eqnarray}\label{eq:pr33a21}
w^\nu\partial_\nu(v^\mu\phi_{,\mu})-v^\nu\partial_\nu(w^\mu\phi_{,\mu})=
\nonumber\\
=(w^\nu v^\mu_{,\nu}-v^\nu w^\mu_{,\nu})\phi_{,\mu}=
\nonumber\\
=w^\mu q_{,\mu}-2m\cos(2\phi)w^\mu\phi_{,\mu}-v^\mu p_{,\mu}=
\nonumber\\
=w^\mu q_{,\mu}-2m p\cos(2\phi)-v^\mu p_{,\mu}.
\end{eqnarray}
Similarly, let us apply the commutator of Lie derivatives $u^\mu \partial_\mu$ and  $w^\mu \partial_\mu$ to $\phi$:
\begin{eqnarray}\label{eq:pr33a22}
w^\nu\partial_\nu(u^\mu\phi_{,\mu})-u^\nu\partial_\nu(w^\mu\phi_{,\mu})=
\nonumber\\
=(w^\nu u^\mu_{,\nu}-u^\nu w^\mu_{,\nu})\phi_{,\mu}=
\nonumber\\
=w^\mu r_{,\mu}-2m\sin(2\phi)w^\mu\phi_{,\mu}-u^\mu p_{,\mu}=
\nonumber\\
=w^\mu r_{,\mu}-2m p\sin(2\phi)-u^\mu p_{,\mu}.
\end{eqnarray}
Let us define vectors $t$ and $s$ by their coordinates:
\begin{eqnarray}\label{eq:pr33a23}
t^\mu=w^\nu v^\mu_{,\nu}-v^\nu w^\mu_{,\nu},
\nonumber\\
s^\mu=w^\nu u^\mu_{,\nu}-u^\nu w^\mu_{,\nu},
\end{eqnarray}
then
\begin{eqnarray}\label{eq:pr33a24}
t^\mu\phi_{,\mu}=w^\mu q_{,\mu}-2m p\cos(2\phi)-v^\mu p_{,\mu},
\nonumber\\
s^\mu\phi_{,\mu}=w^\mu r_{,\mu}-2m p\sin(2\phi)-u^\mu p_{,\mu}.
\end{eqnarray}
One can see that not only $v^0=u^0=0$, but also $w^0=t^0=s^0=0$. Therefore, if in some point $x$ vectors $v$, $u$, and $w$ are not linearly dependent (and we will show that typically they are not), vectors $t$ and $s$ can be presented in the following form:
\begin{eqnarray}\label{eq:pr33a25}
t=a_1 v+a_2 u+a_3 w,
\nonumber\\
s=b_1 v+b_2 u+b_3 w.
\end{eqnarray}
Substituting Eq.~(\ref{eq:pr33a25}) into Eq.~(\ref{eq:pr33a24}) and using Eq.~(\ref{eq:pr33a14}) and Eq.~(\ref{eq:pr33a19}), we obtain:
\begin{eqnarray}\label{eq:pr33a26}
a_1(q-m\sin(2\phi))+a_2(r+m\cos(2\phi))+a_3 p=
\nonumber\\
=w^\mu q_{,\mu}-2m p\cos(2\phi)-v^\mu p_{,\mu},
\nonumber\\
b_1(q-m\sin(2\phi))+b_2(r+m\cos(2\phi))+b_3 p=
\nonumber\\
=w^\mu r_{,\mu}-2m p\sin(2\phi)-u^\mu p_{,\mu}.
\end{eqnarray}
These equations can be regarded as a system of two linear equations in unknowns $\sin(2\phi)$ and $\cos(2\phi)$. The relevant determinant is
\begin{eqnarray}\label{eq:pr33a27}
\left|\begin{array}{cc}-m a_1 & m a_2+2 m p\\
-m b_1+2 m p&m b_2\\
 \end{array}\right|.
\end{eqnarray}
If this determinant does not vanish, this system uniquely defines $\sin(2\phi)$ and $\cos(2\phi)$. We can expect that the sum of their squares will equal unity, as we assumed that vector $B^\mu$ and Majorana spinor $\Phi$ satisfy the system of equations Eqs.~(\ref{eq:pr30},\ref{eq:pr31},\ref{eq:pr32},\ref{eq:pr33a1}). Thus, $2\phi$ is defined uniquely up to a  term $2\pi n$ (n is integer), so $\phi$ is defined uniquely up to a term $\pi n$, and $\Phi$ (see Eq.~(\ref{eq:pr33a7})) is defined uniquely up to a sign, which we can choose arbitrarily in one point and determine by continuity in adjacent points. In the specific example below, vectors $v$, $u$, and $w$ are not linearly dependent, and the determinant does not vanish, so it is reasonable to expect that this is typically the case.
So let us finally prove that typically (in a transversal case), if $B^\mu$ and their temporal derivatives up to the third order ($\dot{B^\mu}$, $\ddot{B^\mu}$, and $B^{\mu(III)}$) are known on the hyperplane defined by a certain value of $x^0$, the temporal derivatives of the fourth order $B^{\mu(IV)}$ can be found from the system for the same hyperplane, thus, the Cauchy problem can be posed and solved for all $x^0$, and therefore the electromagnetic potential $B^\mu$ evolves independently. Indeed, current $J^\mu=\bar{\Phi}\gamma^\mu\Phi$ is a function of $B^\mu$, their temporal derivatives up to the second order, and spatial derivatives of these values (see Eqs.~(\ref{eq:pr30},\ref{eq:pr31})). Therefore, the same is true for spinor $\psi$ (see Eq.~(\ref{eq:pr33a4})) and vectors $v$ and $u$ (see Eq.~(\ref{eq:pr33a13})). Scalars $q$ and $r$ are functions of $B^\mu$, their temporal derivatives up to the third order, and spatial derivatives of these values (see Eq.~(\ref{eq:pr33a13})). The same is true for vector $w$ (see Eq.~(\ref{eq:pr33a17})), scalar $p$ (Eq.~(\ref{eq:pr33a20})), vectors $t$ and $s$ (Eq.~(\ref{eq:pr33a23})), values $a1$, $a2$, $a3$, $b1$, $b2$, $b3$ (Eq.~(\ref{eq:pr33a25})), $\phi$ (Eq.~(\ref{eq:pr33a26})), and, finally, $\Phi$ (Eq.~(\ref{eq:pr33a7})) and all temporal derivatives of $\Phi$ (due to Eq.~(\ref{eq:pr32})) and current $J^\mu=\bar{\Phi}\gamma^\mu\Phi$. Therefore, we can differentiate Eq.~(\ref{eq:pr30}) for $\mu=1,2,3$ twice with respect to $x^0$ and express $B^{\mu(IV)}$ for $\mu=1,2,3$ as functions of $B^\mu$, their temporal derivatives up to the third order, and spatial derivatives of these values. We can also differentiate equation $B^\mu J_\mu=0$ four times with respect to $x_0$ and express $B^{0(IV)}$ as a function of the same values. Thus, the statement is proven.
Let us construct a specific example as an illustration and as a proof that the above does describe a typical situation. It was proven above that for each Majorana spinor $\Phi$ satisfying the free Dirac equation, there exists such electromagnetic potential $B^\mu$ that $\Phi$ and $B^\mu$ satisfy the system of Eqs.~(\ref{eq:pr30},\ref{eq:pr31},\ref{eq:pr32},\ref{eq:pr33a1}). To define such Majorana spinor, it is sufficient to define it at a hyperplane defined by a certain value of $x^0$. The spinor for all values of $x^0$ can then be calculated using the Dirac equation. Moreover, it is not difficult to check that if vectors $v$ and $u$ in a certain point satisfy conditions $v^\mu v_\mu=u^\mu u_\mu=-1$ and $v^0=u^0=v^\mu u_\mu=0$, they define, up to a real scalar factor, such a Majorana spinor $\Phi$ in that point that $v$, $u$, and $\Phi$ satisfy the following equations:
\begin{eqnarray}\label{eq:pr33a28}
v^\mu=\frac{\imath\bar{\Phi}\gamma^0\gamma^\mu\Phi}{\bar{\Phi}\gamma^0\Phi},
\nonumber\\
u^\mu=\frac{\bar{\Phi}\gamma^0\gamma^5\gamma^\mu\Phi}{\bar{\Phi}\gamma^0\Phi}
\end{eqnarray}
(cf. Eq.~(\ref{eq:pr33a13})). Therefore, it is sufficient to define $v$ and $u$ on the hyperplane $x^0=\mathrm{const}$. As a result, the spinor will be defined at that hyperplane up to a factor (a real function), which can be chosen arbitrarily. Let us define $v$ and $u$ for $x^0=0$ as follows:
\begin{eqnarray}\label{eq:pr33a29}
v=(0,v^1,x^2,x^3),
\nonumber\\
u=\left(0,\frac{-x^1 x^3-u^2 x^2}{v^1},u^2,x^1\right)
\end{eqnarray}
where
\begin{eqnarray}\label{eq:pr33a30}
v^1=\sqrt{1-(x^2)^2-(x^3)^2},
\end{eqnarray}
and
\begin{widetext}
\begin{eqnarray}\label{eq:pr33a31}
u^2=\frac{-x^1 x^2 x^3+\sqrt{(x^1 x^2 x^3)^2+(1-(x^3)^2)((1-(x^1)^2)(1-(x^2)^2)
-(x^3)^2)}}{1-(x^3)^2}.
\end{eqnarray}
\end{widetext}
As spinor $\Phi$ satisfies the free Dirac equation, we obtain from Eqs.~(\ref{eq:pr33a13},\ref{eq:pr33a20}):
\begin{eqnarray}\label{eq:pr33a32}
q=0,
\nonumber\\
r=-m,
\nonumber\\
p=0.
\end{eqnarray}
The expressions for $w$, $s$, and $t$ for this example were calculated using Eqs.~(\ref{eq:pr33a17},\ref{eq:pr33a23}), but they are too cumbersome to display here. Furthermore, we only need them (and expressions for $v$ and $u$) for one point, e.g., that with coordinates $x^0=x^1=x^2=x^3=0$:
\begin{eqnarray}\label{eq:pr33a33}
v(0,0,0,0)=(0,1,0,0),
\nonumber\\
u(0,0,0,0)=(0,0,1,0),
\nonumber\\
w(0,0,0,0)=(0,0,1-2 m,-1),
\nonumber\\
t(0,0,0,0)=(0,0,-2 m,-2+2 m),
\nonumber\\
s(0,0,0,0)=(0,-1,0,0).
\end{eqnarray}
Obviously, vectors $v$, $u$, and $w$ are not linearly dependent in this point, so vectors $t$ and $s$ can be presented in the form of Eq.~(\ref{eq:pr33a25}), where
\begin{eqnarray}\label{eq:pr33a34}
a_1=0,
\nonumber\\
a_2=-2+4 m-4 m^2,
\nonumber\\
a_3=2-2 m,
\nonumber\\
b_1=-1,
\nonumber\\
b_2=0,
\nonumber\\
b_3=0.
\end{eqnarray}
Therefore, the determinant of Eq.~(\ref{eq:pr33a27}) equals $m^2(2-4 m+4 m^2)$ and does not vanish (if $m\neq 0$).

 \bibliography{maj2}

\begin{thebibliography}{2}
\expandafter\ifx\csname natexlab\endcsname\relax\def\natexlab#1{#1}\fi
\expandafter\ifx\csname bibnamefont\endcsname\relax
  \def\bibnamefont#1{#1}\fi
\expandafter\ifx\csname bibfnamefont\endcsname\relax
  \def\bibfnamefont#1{#1}\fi
\expandafter\ifx\csname citenamefont\endcsname\relax
  \def\citenamefont#1{#1}\fi
\expandafter\ifx\csname url\endcsname\relax
  \def\url#1{\texttt{#1}}\fi
\expandafter\ifx\csname urlprefix\endcsname\relax\def\urlprefix{URL }\fi
\providecommand{\bibinfo}[2]{#2}
\providecommand{\eprint}[2][]{\url{#2}}

\bibitem[{\citenamefont{Akhmeteli}()}]{Akhm10}
\bibinfo{author}{\bibfnamefont{A.~M.} \bibnamefont{Akhmeteli}},
  \eprint{quant-ph/0509044}.

\bibitem[{\citenamefont{Itzykson and Zuber}(1980)}]{Itzykson}
\bibinfo{author}{\bibfnamefont{C.}~\bibnamefont{Itzykson}} \bibnamefont{and}
  \bibinfo{author}{\bibfnamefont{J.-B.} \bibnamefont{Zuber}},
  \emph{\bibinfo{title}{{Quantum field theory}}}
  (\bibinfo{publisher}{McGraw-Hill}, \bibinfo{year}{1980}).

\end{thebibliography}

\end{document}